# Collective Coordinate Models of Domain Wall Motion in Perpendicularly Magnetized Systems under the Spin Hall Effect and Longitudinal Fields

S. Ali Nasseri[1,2], Simone Moretti[3], Eduardo Martinez[3], Claudio Serpico[1,4], Gianfranco Durin[1,5]

[1] ISI Foundation – Via Alassio 11/c – 10126 Torino, Italy
[2] Politecnico di Torino - Corso Duca degli Abruzzi 24, 10129 Torino, Italy
[3] University of Salamanca - Cardenal Plá y Deniel, 22, 37008 Salamanca, Spain
[4] University of Naples Federico II - Via Claudio 21, 80125 Napoli, Italy
[5] Istituto Nazionale di Ricerca Metrologica (INRIM) - Strada delle Cacce 91, 10135 Torino, Italy
ali.nasseri@isi.it

*Abstract* —Recent studies on heterostructures of ultrathin ferromagnets sandwiched between a heavy metal layer and an oxide have highlighted the importance of spin-orbit coupling (SOC) and broken inversion symmetry in domain wall (DW) motion. Specifically, chiral DWs are stabilized in these systems due to the Dzyaloshinskii-Moriya interaction (DMI). SOC can also lead to enhanced current induced DW motion, with the spin Hall effect (SHE) suggested as the dominant mechanism for this observation. The efficiency of SHE driven DW motion depends on the internal magnetic structure of the DW, which could be controlled using externally applied longitudinal in-plane fields. In this work, micromagnetic simulations and collective coordinate models are used to study current-driven DW motion under longitudinal in-plane fields in perpendicularly magnetized samples with strong DMI. Several extended collective coordinate models are developed to reproduce the micromagnetic results. While these extended models show improvements over traditional models of this kind, there are still discrepancies between them and micromagnetic simulations which require further work.

*Index: magnetic DW motion – PMA material – spin Hall effect (SHE)*

## I. INTRODUCTION

Manipulating magnetic domain walls (DWs) within nanostructures has been linked with applications in the development of spintronic logic [1-4], storage [5-13] and sensing devices [14]. Devices based on this technology benefit from low power dissipation, non-volatile data retention, radiation hardness, faster manipulation of data, high areal densities and absence of mechanical parts. The potential applications of DW based devices have led to increased interest within the scientific community in developing models which can qualitatively or quantitatively describe DW motion under applied fields and currents.

Recent studies on DW motion have focused on heterostructures made of a ferromagnetic layer sandwiched between two heavy metal layers or a heavy metal layer and an oxide layer [15-17]. The importance of the Dzyaloshinskii-Moriya interaction (DMI) in DW motion in such systems has recently been highlighted [18-19]. In the case of current driven DW motion, interfacial induced torques due to the spin Hall and Rashba effects have been shown to be present in addition to the spin transfer torque mechanism [20-23]. These features along with the higher DW velocities achieved in these systems have rendered current driven DW motion in ferromagnetic heterostructures interesting both from a fundamental perspective and for applications.

In perpendicularly magnetized heterostructures with DMI, applied fields in-plane of the sample could be used to control DW chirality and the direction of DW motion [24-29]. Micromagnetic ($\mu M$) simulations of such systems are in agreement with experiments [22], showing an increase in DW velocity with fields parallel to the internal magnetization of the DW. However, the conventional collective coordinate models (q-$\Phi$ and q-$\Phi$-$\chi$) fail to reproduce these results [25, 28]. This calls for improvements in analytical modeling of DW motion in such systems.

In this paper, DWs driven by the spin Hall effect (SHE) under longitudinal in-plane fields are studied in perpendicular magnetocrystalline anisotropy (PMA) materials. In order to improve the agreement of collective coordinate models (CCMs) with *micromagnetic* simulations, the conventional tilted CCM is extended by including the DW width, and canting of the magnetization in the domains. This extended CCM shows qualitative improvements in predicting DW motion over a larger range of longitudinal fields.

## II. THE LLG EQUATION

As a case study, current-driven DW motion along a Pt/CoFe/MgO nanowire is evaluated in this work. The Landau-Lifshitz-Gilbert (LLG) equation for such a ferromagnetic heterostructure reads:

$$\frac{d\vec{m}}{dt} = -\gamma_0 \vec{m} \times \vec{H}_{eff} + \alpha \vec{m} \times \frac{d\vec{m}}{dt} - \gamma_0 H_{SL} \vec{m} \times (\vec{m} \times \hat{u}_y) \qquad (1)$$

where the effective field is $H_{eff} = -\frac{1}{\mu_0 M_s}\frac{\delta E}{\delta \vec{m}}$ and the energy density of the system is written as:

$$e_D = \frac{dE}{dV} = A \sum |\nabla \vec{m}_i|^2 + K_U \sin^2\theta - \mu_0 M_s \vec{m} \cdot \vec{H}_a - \frac{1}{2}\mu_0 M_s \vec{m} \cdot \vec{H}_d + D[m_z \nabla \cdot \vec{m} - (\vec{m} \cdot \nabla)m_z] \qquad (2)$$

The energy density of the system includes contributions from exchange, uniaxial magnetocrystalline anisotropy, magnetostatics,

DMI and applied fields. The last term in eq. (1) is the Slonczewski-like torque due to the SHE, which is characterized by $H_{SL} = \frac{\hbar \theta_{SHE} J}{2 \mu_0 e M_s t_f}$, where $J$ is the current, $\theta_{SHE}$ is the SHE angle and $t_f$ denotes the thickness of the ferromagnetic layer [30]. Note that the effect of Spin Transfer Torques (STTs) has been neglected in this study, due to the small thickness of the ferromagnetic layer [23, 26, 28]. Moreover, it was assumed that only the Slonczewski-like torques arising from SHE give rise to steady DW motion; an assumption which has been supported by other studies [26, 30].

All micromagnetic simulations in this work were performed using the mumax$^3$ package [31]. The dimensions of the CoFe strip used in this study are 2.8 μm x 160 nm x 0.6 nm. Typical parameters for the material stack were adopted [26, 28]: saturation magnetization $M_s$ = 700 kA/m, exchange constant A = 0.1 pJ/m, uniaxial perpendicular anisotropy constant $K_u$ = 480 kJ/m$^3$, Gilbert damping α = 0.3, DMI strength D = -1.2 mJ/m$^2$, and SHE angle $\theta_{SH}$=0.07.

### III. DW STRUCTURE UNDER IN-PLANE FIELDS

To better understand the effect of longitudinal in-plane fields on DW structure, micromagnetic simulations were conducted both on a static (non-moving) DW and a moving DW under the application of longitudinal in-plane fields in the range -225 mT < $B_x$ < 325 mT. Figure 1 illustrates the results of this study.

The geometric tilting of the DW during motion is clearly seen in Figure 1.(b). The application of $B_x$ tilts the magnetization in the domains into the plane, reducing the $m_z$ component and $\theta = \mathrm{acos}(m_z)$. When the DMI and $B_x$ are supporting each other within the DW ($B_x$ > 0 in Figure 1), the DW width increases and the DW is further stabilized. In cases where the DMI and $B_x$ are competing ($B_x$ < 0 in Figure 1), a sufficiently large in-plane field can change the chirality of the DW and, in the static case, tilt the DW plane.

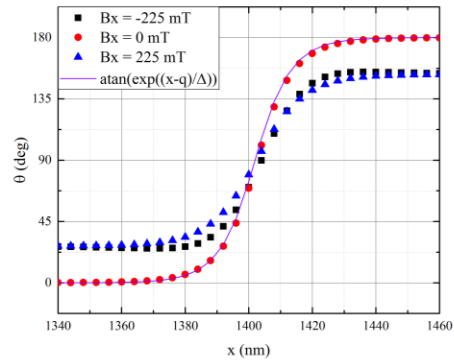

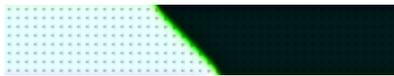
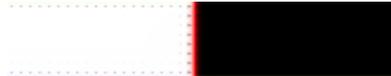
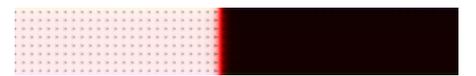

$B_x$ = -225 mT ····· $B_x$ = 0 ····· $B_x$ = +225 mT

*(a) Static structure of the DW under longitudinal in-plane fields and snapshots of the DW.*

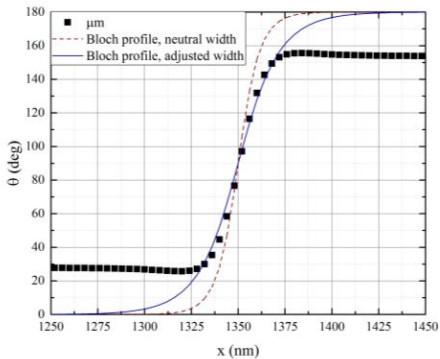
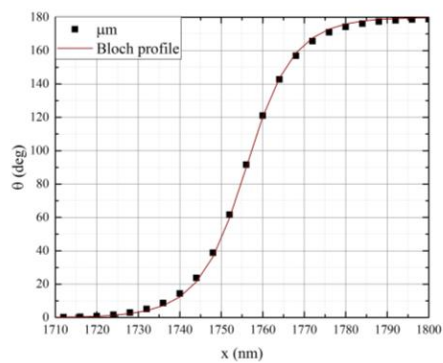
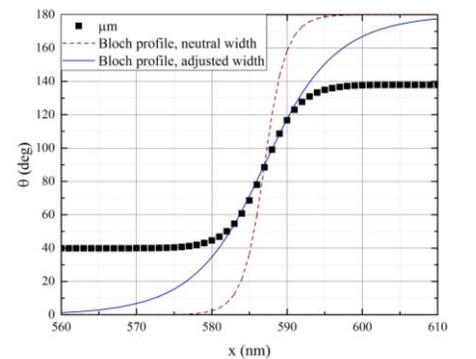

$B_x$ = -225 mT ····· $B_x$ = 0 ····· $B_x$ = +325 mT

*(b) DW Structure and snapshots of the moving DW under the application of longitudinal fields 10ns after the start of motion under the application of a current density of 0.1 TA/m2 (the arrow shows direction of motion.)*

Figure 1. DW structure under (a) static and (b) dynamic conditions. The DW structure was characterized by $\theta = \mathrm{acos}(m_z)$ component of the magnetization, extracted from the center of the nanowire. The neutral width is the width of the DW when no in-plane fields are applied and was calculated using $\Delta = \sqrt{\frac{A}{K_u - 0.5 \mu_0 M_s^2 N_z}}$. It is clear that the DW maintains its shape in motion, and the static and dynamic DW follow the Bloch profile in the transition region between the two domains.

Figure 1 shows that in the absence of in-plane fields, the Bloch profile ( $\theta(x,y,t) = 2\mathrm{atan}\left(\exp\left(\frac{x-q}{\Delta}\right)\right)$ ) describes the change in $\theta$ acceptably both for a static and moving DW, while the DW profile slightly deviates from the Bloch profile under longitudinal fields. It can be shown that the Bloch profile can still fit the transition from one domain to the next under longitudinal in-plane fields, if the value of the DW width is adjusted or a prefactor is added to the ansatz.

The simulations in Figure 1 illustrate two important effects of longitudinal in-plane fields. While the transition between the domains (and the internal structure of the DW) is almost unaffected by the applied in-plane fields, the DW width changes (in fact, for Bx > 225 mT, it reaches 3-4 times its value when no fields are applied). More importantly, the magnetization in the domains is canted; this means that instead of $\theta$ = 0 or 180 degrees in the domain, 0 < $\theta$ < 180 degrees in the domains. Both of these features have important consequences in developing CCMs for magnetic DW motion as discussed later.

### IV. MICROMAGNETIC RESULTS FOR DW MOTION

Figure 2 illustrates the results of micromagnetic simulations of SHE driven DW motion under an applied current of 1 TA/m². The velocity curve as a function of longitudinal fields possesses a point of inflection around $B_x = 0$, a characteristic feature which could be used to assess whether collective coordinate models are predicting the right trends in velocity. The DW width has a minimum at $B_x$ = -100 mT which corresponds to the field at which the DW tilts under static conditions. While the application of negative longitudinal fields tends to lead to changes in both magnetization angle at the center of the DW and the tilting angle of the DW, in the case of positive fields the magnetization angle in the DW stays constant for $B_x$>100 mT. This fixing of the magnetization angle inside the DW is understandable, as the positive in-plane field would try to align the magnetic moments inside the DW with itself and stabilize the DW.

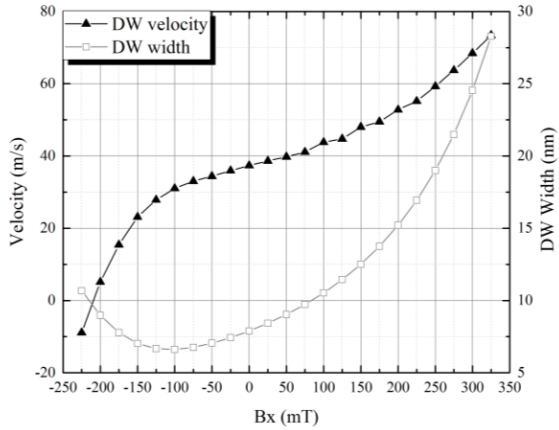

*(a) DW velocity and width.*

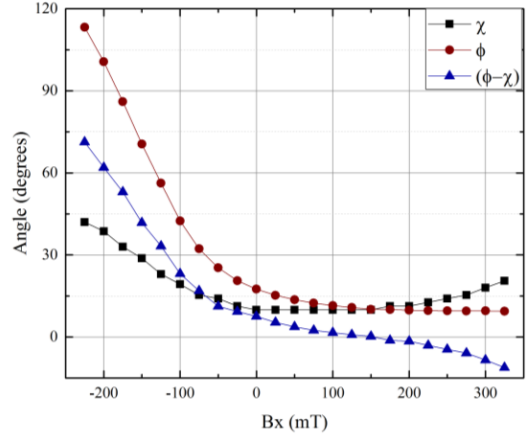

*(b) Magnetization angle at the center of the DW ($\Phi$), tilting angle of the DW ($\chi$) and their difference ($\Phi$-$\chi$).*

Figure 2. Variation of different domain wall properties with in-plane fields as calculated from micromagnetic simulations on a Pt/CoFe/MgO system under the application of a current density of 0.1 TA/m². The DW velocity has a point of inflection at $B_x$ = 0 mT. Moreover, a critical longitudinal field exists for which the DW velocity is zero.

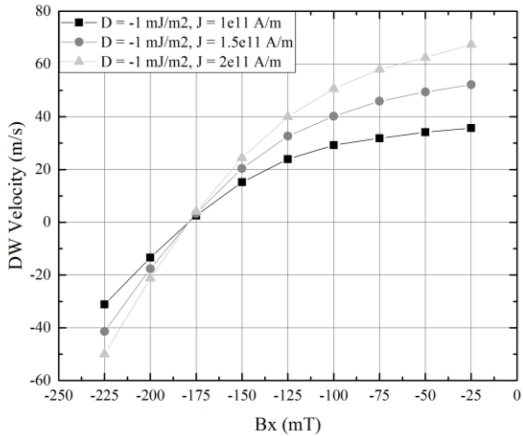

*(a) Effect of current density variation.*

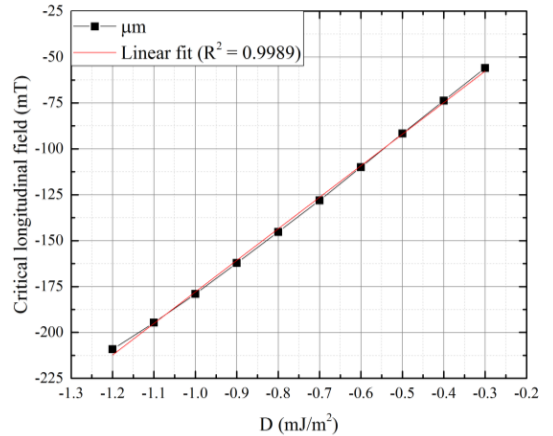

*(b) Variation of critical longitudinal field with DMI strength.*

Figure 3. Effect of current density and DMI strength on the critical longitudinal in-plane field for zero DW velocity. It is clear that (a) the longitudinal field inducing zero DW velocity is independent of the current density, and (b) it has a linear relationship with the DMI strength.

The nonlinear behaviour of velocity against $B_x$ seen in Figure 2 may seem to contradict experimental results at first, as published experimental results show a linear behaviour [33-35]. Upon further analysis, we found that in most experiments, the range of longitudinal fields used was restricted to |Bx|<100. Within this range of fields, our results also show a somewhat linear behaviour. Observance of a nonlinear behaviour similar to what is presented here, depends on the material stack parameters, and the range of in-plane fields used.

One of the intriguing features of the application of longitudinal fields to SHE driven DW motion is that a longitudinal field exists at which the direction of DW motion reverses. According to Figure 2, a zero velocity corresponds to -225mT < Bx < -200mT with $\Phi - \chi \sim$ 60-70 degrees. To better understand the relationship between the DMI strength (D) and the longitudinal field at which DW velocity is zero, more simulations were performed for different values of DMI strength and applied current. The results of this study, depicted in Figure 3, show that this critical longitudinal field is almost independent of the current density and has a linear dependence on the strength of the DMI. As such, DW motion in nanowires under longitudinal in-plane fields could be used to measure DMI strengths for |D| < 1.2 mJ/m² in samples with similar material properties to Pt/CoFe/MgO. DMI strengths higher than 1.2 mJ/m² are harder to study in nanowires, as much higher longitudinal fields will be required (such high fields could not be simulated).

## V. Collective Coordinate models

Based on the LLG equation and using a Lagrangian description, a CCM was developed taking into account four time dependent collective coordinates, namely the DW position (q), the magnetization angle at the center of the DW ($\Phi$), the DW width ($\Delta$), and the tilting angle of the DW plane ($\chi$). Figure 4 shows the spherical and collective coordinates used in deriving the CCMs.

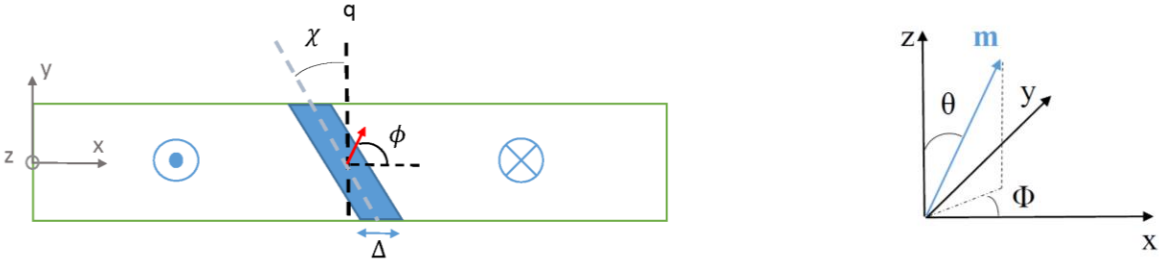

*(a) Top view of the nanowire, showing the coordinates q, Δ and χ.*    *(b) The spherical coordinates used in this work.*
*Figure 4. The coordinate systems used in deriving the analytical description.*

An adjusted ansatz based on a tilted Bloch profile [25] with the addition of two prefactors was used to connect the collective coordinates q, Δ and χ with the spherical coordinate θ:

$$\theta(x,y,t) = 2P_1 \text{atan}\left(\exp\left(\frac{(x-q)\cos\chi + y\sin\chi}{P_2\Delta}\right)\right) \quad (3)$$

Where x is the position along the length of the nanowire, y is the position along the width of the wire, $P_1$ is an ansatz prefactor and $P_2$ is a prefactor for the DW width. $P_1$ and $P_2$ are assumed to be only functions of the applied longitudinal fields and were extracted from micromagnetic simulations to try to tune the ansatz to better predict the micromagnetic simulations. We assumed that the chirality preferred by the DW leads to magnetization pointing along the x-direction (which is valid for D<0); the definition of the collective coordinate $\Phi$ needs to be adjusted to $\Phi+\pi$ to model the D>0 case. Moreover, we also assumed that the left domain is pointing up; a negative should be added to the $P_2$ factor to model cases where the left domain is pointing down.

The effect of canting was also included in deriving the CCMs. While traditionally, the energy densities used in deriving SSMs are integrated from 0 to $\pi$ along the ansatz, the integration in this work was done from $\theta_c$ to $\pi - \theta_c$ (where $\theta_c$ is the canting angle in the domains) to take into account that the ansatz is only valid in the transition region between the two domains, as highlighted in Figure 1. In the domains, under the application of longitudinal in-plane fields and far away from the edges of the system, the magnitude of the canting angle may be calculated using $\theta_c = \text{asin}\left(\frac{M_s B_x}{2K_u + \mu_0 M_s^2(N_x - N_z)}\right)$ which is derived from energy minimization ($\frac{\partial E}{\partial \theta} = 0$). As depicted in Figure 5, this formulation predicts the canting in the domains perfectly for the range of longitudinal in-plane fields under study.

To derive the CCMs, the relevant energy densities were integrated using the ansatz with limits set based on the canting angle. For the system being studied, the four coordinate CCM has the following implicit form:

$$\dot{\phi} + \alpha P_1 \cos\theta_c \frac{\dot{q}}{P_2\Delta}\cos\chi = \frac{\pi - 2\theta_c}{2} P_1 \mu_0 \gamma H_{SL} \cos\phi \quad (4)$$

$$P_1 \frac{\dot{q}}{P_2\Delta}\cos\chi - \alpha\cos\theta_c \dot{\phi} = \frac{1}{2}\mu_0\gamma\cos\theta_c M_s(N_y - N_x)\sin 2(\phi - \chi) + \frac{\pi - 2\theta_c}{2}\mu_0\gamma\left[H_x \sin\phi - \frac{D}{\mu_0 M_s P_2\Delta}\sin(\phi - \chi)\right] \quad (5)$$

$$\frac{\alpha B P_1^2}{2}\left(\frac{\dot{\Delta}}{P_2\Delta}+\frac{\dot{\chi}}{\cos\chi}\sin\chi\right)=\frac{\gamma}{M_s}\cos\theta_c\left[\frac{A}{(P_2\Delta)^2}-P_2K\right]+\frac{\pi-2\theta_c}{2}\mu_0\gamma H_x\cos\phi \tag{6}$$

$$-\frac{\alpha B P_1^2}{2}\left(\frac{\dot{\Delta}}{P_2\Delta}\sin\chi+\frac{\dot{\chi}}{\cos\chi}\left[\frac{1}{6B}\left(\frac{w}{P_2\Delta}\right)^2+\sin^2\chi\right]\right)=\frac{\pi-2\theta_c}{2}\mu_0\gamma\left[P_1\frac{D}{\mu_0 M_s P_2\Delta}\sin\phi-H_x\cos\phi\right] \\ +\frac{\gamma}{M_s}\cos\theta_c\sin\chi\left[\frac{A}{(P_2\Delta)^2}+K-\frac{1}{2}\mu_0 M_s^2(N_y-N_x)\cot\chi\sin 2(\phi-\chi)\right] \tag{7}$$

where $K=K_u+\frac{1}{2}\mu_0 M_s^2\left(N_x\cos^2(\phi-\chi)+N_y\sin^2(\phi-\chi)-N_z\right)$ and w is the width of the nanowire. The demagnetizing factors were estimated using the approach proposed by Aharoni [35].

The integration constant $0<B\leq\frac{\pi^2}{6}$ is a parameter dependent on the canting angle and has the following form:

$$\begin{aligned}B=&\,2\left[Li_2\left(-\cos\left(\frac{\theta_c}{2}\right)\right)-Li_2\left(-\sin\left(\frac{\theta_c}{2}\right)\right)-Li_2\left(1-\cos\left(\frac{\theta_c}{2}\right)\right)+Li_2\left(1-\sin\left(\frac{\theta_c}{2}\right)\right)\right]\\ &-2\cos^2\left(\frac{\theta_c}{2}\right)\left[1-\log\left(\cos\left(\frac{\theta_c}{2}\right)\right)\right]+2\sin^2\left(\frac{\theta_c}{2}\right)\left[1-\log\left(\sin\left(\frac{\theta_c}{2}\right)\right)\right]\\ &-\cos^2\left(\frac{\theta_c}{2}\right)\log\left(\sin^2\left(\frac{\theta_c}{2}\right)\right)\left[2\log\left(\cos\left(\frac{\theta_c}{2}\right)\right)-1\right]+\sin^2\left(\frac{\theta_c}{2}\right)\log\left(\cos^2\left(\frac{\theta_c}{2}\right)\right)\left[2\log\left(\sin\left(\frac{\theta_c}{2}\right)\right)-1\right]\\ &-\log\left(\sin^2\left(\frac{\theta_c}{2}\right)\right)+\log\left(\cos^2\left(\frac{\theta_c}{2}\right)\right)+2\log\left(\cos\left(\frac{\theta_c}{2}\right)\right)\log\left(1+\cos\left(\frac{\theta_c}{2}\right)\right)-2\log\left(\sin\left(\frac{\theta_c}{2}\right)\right)\log\left(1+\sin\left(\frac{\theta_c}{2}\right)\right)\\ &+2\left[\cos\theta_c+(\cos\theta_c+1)\ln^2\cos\frac{\theta_c}{2}+(\cos\theta_c-1)\ln^2\sin\frac{\theta_c}{2}\right]\end{aligned} \tag{8}$$

In eq. (8), $Li_2$ is the polylogarithm function. The polylogarithmic part of B may be estimated with good accuracy using a series expansion.

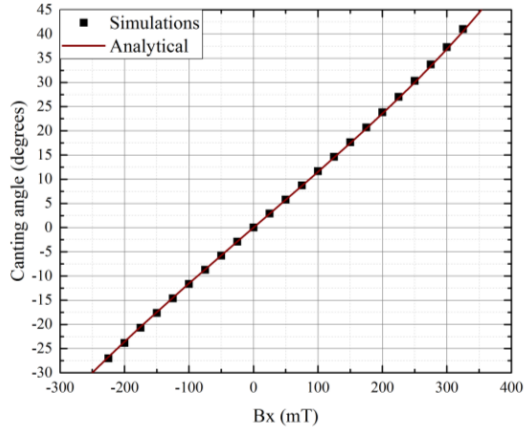

*Figure 5. Comparison of the analytical prediction of the canting angle to micromagnetic results. The figure clearly shows a perfect match between analytical and micromagnetic results.*

Note that, the same approach may be used to derive a two coordinate q-Φ model, and the three coordinate q-Φ-χ and q-Φ-Δ models including the effect of canting. While equations 5 and 6 stay the same in all models, the third equation in the three coordinate models need to be re-derived and has different forms compared to equations (6) and (7) above. The traditional CCMs [25, 36-37] may be derived from equations (4-7) by setting $P_1=P_2=1$ and $\theta_c=0$, taking into account the relevant coordinates in each case.

Figure 6 illustrates the steady state predictions of DW velocity ($\dot{q}$), magnetization angle (Φ), tilting angle (χ), and DW width (Δ) from micromagnetic simulations compared to different forms of the CCMs. Comparing CCMs without canting and prefactors to the micromagnetic simulations (first column of images in Figure 6) shows that all four coordinates are necessary to be able to properly model the system. This is clear, as only the model with four coordinates predicts a point of inflection for the velocity curve at $B_x=0$ similar to the micromagnetic simulations, and can qualitatively predict the right trends for the collective coordinates for positive and negative fields. Overall, none of these models are able to accurately predict DW velocity when longitudinal fields are applied, and their predictions are only accurate in the absence of longitudinal fields.

The addition of canting in the derivation of the collective coordinate models improves the accuracy of predictions as depicted in Figure 6. For the cases with negative in-plane fields (which gives rise to the tilting of the DW under static conditions), models including all four coordinates consistently predicted the velocity accurately, but could not be integrated for fields $B_x<-150$ mT. For positive in-plane fields, the q-Φ and q-Φ-χ models reproduced the results up to $B_x=50$ mT and fail for higher fields, while models including the DW width are able to reproduce a curvature opposite that for negative fields, albeit diverging. Unfortunately, non of the models are able to predict the zero-velocity crossing point, which would be of interest for predicting DMI strength.

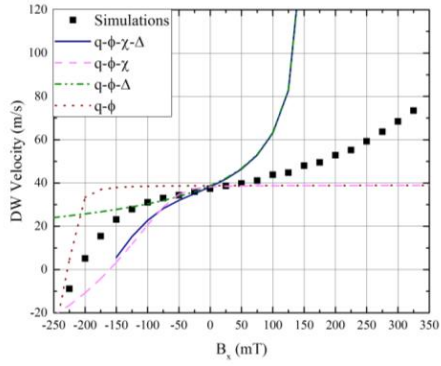
*Analytical models with no canting or prefactors.*

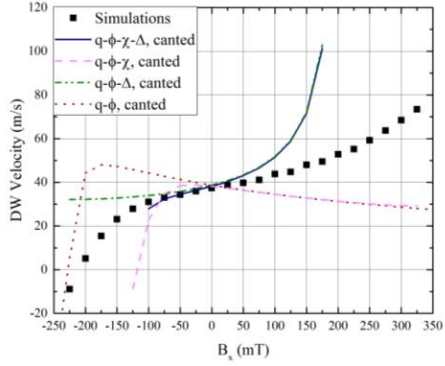
*Analytical models with canting included.*
*(a) DW velocity (q̇).*

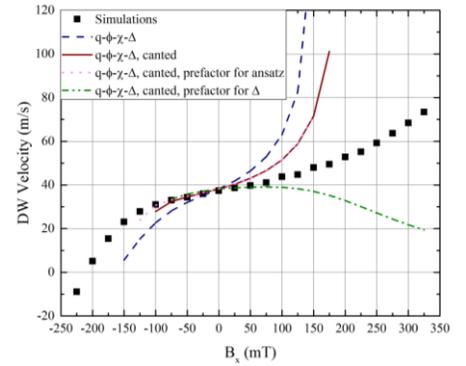
*Different forms of the q-Φ-χ-Δ model.*

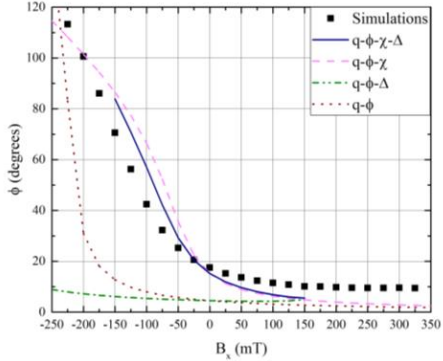
*Analytical models with no canting or prefactors.*

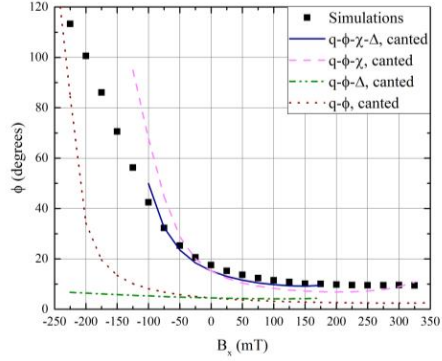
*Analytical models with canting included.*
*(b) Magnetization angle of DW (Φ).*

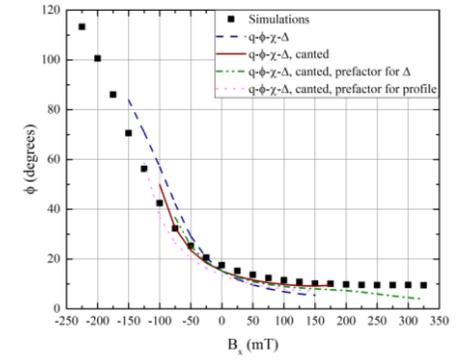
*Different forms of the q-Φ-χ-Δ model.*

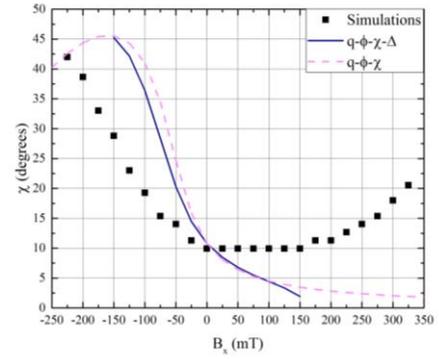
*Analytical models with no canting or prefactors.*

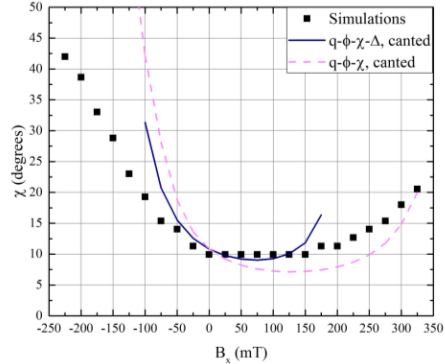
*Analytical models with canting included.*
*(c) Tilting angle of DW (χ).*

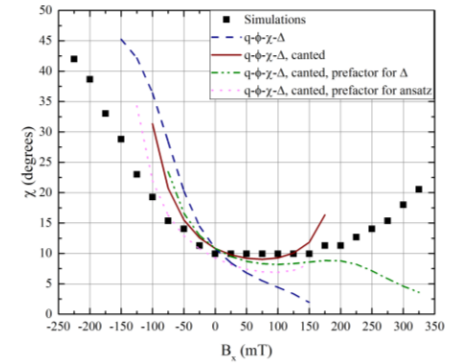
*Different forms of the q-Φ-χ-Δ model.*

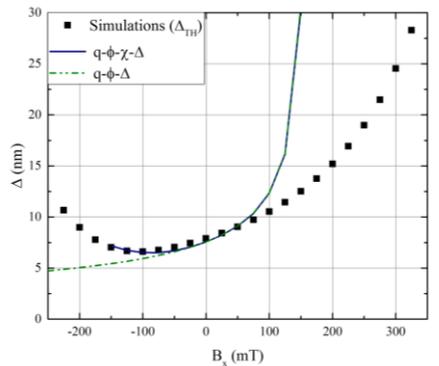
*Analytical models with no canting or prefactors.*

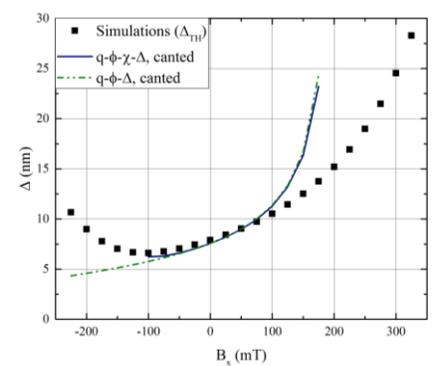
*Analytical models with canting included.*
*(d) DW Width (Δ).*

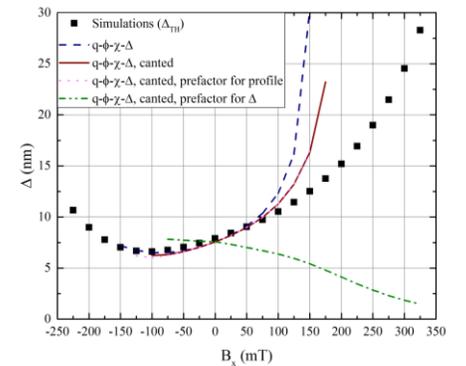
*Different forms of the q-Φ-χ-Δ model.*

*Figure 6. Predictions from analytical models compared to micromagnetic simulations. Clearly, only the four coordinate models are able to reproduce the characteristic change in the curvature of the DW velocity curve. Addition of canting improves the accuracy of the models, with an almost exact prediction of velocity for negative longitudinal fields. Corrections to the DW width seem to be of importance in improving predictions for positive applied fields.*

The low accuracy of the CCMs in the case of high positive longitudinal fields could be attributed to two factors. Firstly, as highlighted in Figure 6.d, we observed that the traditional and canted CCMs with DW width as a coordinate miscalculate the DW width for the case of positive longitudinal fields, which would in turn affect DW velocity predictions. This had already been described in previous work [28]. Second, Figure 6 shows that under $B_x > 100$ mT, the models with tilting ($\chi$) tend to calculate the magnetization angle of the DW ($\Phi$) correctly, while they miscalculate the tilting angle (likely due to miscalculation of the DW width). This could explain the inaccuracy of analytical models, as such models only rely on the perturbation of these angle as the major coordinate driving magnetization dynamics.

To further understand the effect of DW width on the dynamics, prefactors were extracted from micromagnetic simulations to match the DW width in the Bloch profile to that of micromagnetic simulations. Two cases were studied:

1. $P_2 = 1$ and $P_1 = \begin{cases} -0.0006\, B_x + 0.966 & B_x > 0 \\ -3.253 \times 10^{-6}\, B_x^2 - 2.116 \times 10^{-4}\, B_x + 0.9619 & B_x < 0 \end{cases}$

2. $P_1 = 1$ and $P_2 = \begin{cases} 6.598 \times 10^{-8}\, B_x^3 - 8.399 \times 10^{-6}\, B_x^2 + 0.003697\, B_x + 0.9857 & B_x > 0 \\ -9.517 \times 10^{-8}\, B_x^3 - 5.309 \times 10^{-6}\, B_x^2 + 0.00205\, B_x + 0.998 & B_x < 0 \end{cases}$

Results of these models are compared to the traditional and canted four coordinate models in the third column of images in Figure 6. The addition of a prefactor to the DW width (case of $P_1 = 1$ above) seems to increase the area of applicability of the models and improve the velocity predictions. Yet, Figure 6.d reveals that despite accuracy in predicting the DW velocity and the angles up to $B_x = 125$ mT, the model with prefactor for DW width is in fact not predicting the correct DW width for $B_x > 25$ mT. Addition of a prefactor to the ansatz as a whole ($P_2 = 1$ above) does not lead to major improvements. It seems that the only approach to resolve these issues is using an inherently canted DW profile which is currently under investigation.

## VI. Conclusion

We studied the motion of DWs driven by the SHE under the application of longitudinal in-plane fields in a Pt/CoFe/MgO system. Our study revealed that the DW maintains its structure under longitudinal in-plane fields; however, the DW width increases and magnetization in the domains become canted. Micromagnetic simulations revealed a critical longitudinal in-plane field at which the DW velocity is zero, independent of the current density applied. This field could be used as a measure of DMI in experiments.

Finally, new CCMs were proposed to characterize the motion of DWs under the conditions studied in this paper. We found that only an CCM with four collective coordinates (namely DW position q, DW width $\Delta$, DW tilting angle $\chi$ and magnetization at the center of the DW $\Phi$) is able to reproduce the characteristic shape of the DW velocity versus longitudinal field curve and predict the right trends for other collective coordinates. The simple q-$\Phi$-$\chi$-$\Delta$ model was extended by inclusion of canting in the domains, which improved model accuracy with a model able to accurately predict DW motion for $-150$ mT $< B_x < 50$ mT. Other approaches to improving the accuracy of the models such as adding prefactors to the ansatz were also studied with limited success.

### Acknowledgements

This study was conducted as part of the Marie Currie ITN WALL project, which has received funding from the European Union's Seventh Framework Programme for research, technological development and demonstration under grant agreement no. 608031.